\documentclass {acmart}

\usepackage{booktabs} % For formal tables
\usepackage{algorithm}
\usepackage{amsmath}
\usepackage[noend]{algpseudocode}

\setcopyright{none}
\begin{document}
\title{Weighted Random Walk Sampling for Multi-Relational Recommendation \footnote{(fvahedia, rburke, mobasher) @cs.depaul.edu}}

\author{Fatemeh Vahedian, Robin Burke \& Bamshad Mobasher}
\affiliation{%
  \institution{Center for Web Intelligence, DePaul University}
  \streetaddress{243 S Wabash ave}
  \city{Chicago} 
  \state{Illinois } 
  \postcode{60604}
}
\email{  (fvahedia, rburke, mobasher) @cs.depaul.edu}

\begin{abstract}
In the information overloaded web, personalized recommender systems are essential tools to help users find most relevant information. The most heavily-used recommendation frameworks assume user interactions that are characterized by a single relation. However, for many tasks, such as recommendation in social networks, user-item interactions must be modeled as a complex network of multiple relations, not only a single relation. Recently research on multi-relational factorization and hybrid recommender models has shown that using extended meta-paths to capture additional information about both users and items in the network can enhance the accuracy of recommendations in such networks. Most of this work is focused on unweighted heterogeneous networks, and to apply these techniques, weighted relations must be simplified into binary ones. However, information associated with weighted edges, such as user ratings, which may be crucial for recommendation, are lost in such binarization. In this paper, we explore a random walk sampling method in which the frequency of edge sampling is a function of edge weight, and apply this generate extended meta-paths in weighted heterogeneous networks. With this sampling technique, we demonstrate improved performance on multiple data sets both in terms of recommendation accuracy and model generation efficiency. 
\end{abstract}



% We no longer use \terms command


\maketitle

\section{Introduction}

\noindent Recommender systems based on complex heterogeneous  networks have been studied extensively in recent years. A heterogeneous information network (HIN) is defined as a network with multiple types of nodes \cite{shi2015survey} (for example, user, groups, pages and photos) and multiple types of edges (for example, a ``friendship'' relation between  users, ``follow'' relation between user and pages and a ``like'' relation between a user and a post).  Such networks are a natural way to express the multiplicity of connections between types of information in social media applications: for example,  users, employers, interest groups, educational institutions, job postings, posts, and comments are different entities in the LinkedIn social network, connected by a variety of relations. The complexity of heterogeneous networks poses two challenges for recommender systems: (1) the problem of integrating a wide variety of data effectively into a recommendation framework, and (2) the problem of responding to many potential recommendation tasks, because of the wide variety of items present.

The main intuition in work with heterogeneous information networks is that nodes with meaningful semantics can be found by following certain typed paths on a heterogeneous network \cite{yu2014personalized}. In other words, in a heterogeneous information network, two nodes can be connected via different types of paths. Due to the multiplicity of node and edge types in heterogeneous networks, these paths may contain different node types and link types in different orders and they can have various lengths. For example, the relation between LinkedIn users $u$ and $v$ might be friend of a friend, friend of co-worker, or friend of a fellow alumnus to name just a few of the possible two-step paths. In heterogeneous network terminology, these different relations composed of sequences of edge types are known as \textit{meta-paths}. \textit{Meta-path expansion} is the process that begins at a starting node and follows all possible paths that conform to the meta-path to yield a set of destination nodes.

Recent research on multi-relational recommendation in heterogeneous networks has shown that extended meta-paths (those that connect beyond the immediate neighbors of users and target items) are effective in generating relations on which to base recommendations. This benefit has been demonstrated with several different recommendation algorithms including  multi-relational matrix factorization \cite{DBLP:conf/recsys/VahedianBM15,DBLP:conf/flairs/VahedianBM16}, weighted hybrid of low-dimensional components \cite{DBLP:conf/um/BurkeVM14,DBLP:conf/recsys/BurkeV13,DBLP:conf/recsys/VahedianB14,DBLP:conf/recsys/Vahedian14} and non-negative matrix factorization \cite{yu2014personalized,Yu:2013:RHI:2507157.2507230}. 

\begin{figure}[tbh]
	\centering
	\includegraphics[width=8cm]{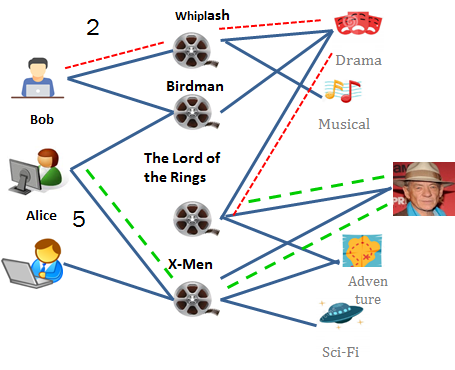}
	\caption{\label{fig:wnet} Example of weighted network in movie dataset}
\end{figure}

However, these models all assume a uniform preference associated with all relations in the network. In many networks, however, there are weighted edges that provide important information: for example, if users provide explicit rating values, these encode useful information about preferences, and it is best not to ignore them. For example, Figure \ref{fig:wnet} represents a fraction of movie dataset in which user provides rating value for each movie in range of $1-5$. In this figure user ``Bob'' rated movie ``Whiplash'' 2 which indicates that this user did not like the movie. A meta-path expansion following $user-movie-genre-movie$ usually takes this edge and adds all ``Drama" movies to the list of ``Bob'' preferences ignoring the rating value. However, the reason that ``Bob'' rated this movie 2 may be that he does not enjoy the ``Drama'' genre. Therefore, following a meta-path $Bob \rightarrow Whiplash \rightarrow Drama \rightarrow *$ can be a poor user preference projection for this user. On the other hand, ``Alice'' rated movie ``Xmen'' 5, which represents the highest interest in that movie. Therefore, a meta-path following $Alice \rightarrow Xmen \rightarrow "Ian McKellen" \rightarrow * $ can be a stronger representation of user preferences based on an actor of movies.

What we see from this example is that the recommender system should give greater preference to paths containing highly-weighted edges over paths containing low-weighted edges. The only previous work on weighted heterogeneous networks breaks each meta-path into various similar rated paths \cite{Shi:2015:SPB:2806416.2806528} then combines all of them together. This method multiplies the problem of meta-path generation (typically the most computationally-expensive step in model generation). This model uses weighted meta-paths only for the calculation of user-user similarities: thus, it works only with meta-paths that start from a user node and end with another user node. The technique cannot be generalized for other type of meta-paths, starting from an arbitrary node type and ending to another node type.

In our prior work~\cite{vahedian2016weighted}, we introduced a random walk sampling method for heterogeneous weighted graphs in which meta-paths are generated using exponential sampling that prefers highly-rated edges, and build a recommendation model from the resulting collection of paths. In this paper, we provide a more detailed explanation of the weighted random walk sampling algorithm and provide experimental results for three real world datasets. We show improvements in both accuracy and efficiency of using weighted meta-path sampling and compare the accuracy of this method with three graph-based recommendation algorithms. 

\section{Random Walk Sampling}	
\noindent Christoffel et al.\cite{Christoffel:2015:BWA:2792838.2800180} introduced a random walk sampling algorithms to calculate the transition probability in a random walk model to rank items and generate recommendation model. This model works from an unweighted bipartite graph which represents the binary relations among user and items. Building on this approach, we propose a method for generating meta-paths in a heterogeneous network using biased random walk sampling. This method has the advantage of creating greater efficiency in meta-path generation and allowing for sensitivity to user ratings.

The goal of meta-path generation is to create a relation based on paths through the network. For example, the extended $user-movie-actor-movie-user$ meta-path enables the system to start with a given user and find other users that have watched movies containing actors in common with the user's movies. The semantics of this operation of meta-path expansion is that the end result is a set of destination nodes (in this case, users) weighted by how many of the expanded meta-paths reach that node. 

A random-walk version of this process chooses edges from the next relation in the meta-path randomly instead of following all possible paths. This is more efficient than generating all paths and the number of samples can be chosen to be large enough to provide a good estimate of what a full expansion would provide~\cite{Christoffel:2015:BWA:2792838.2800180}. In this work, we look specifically at networks involving a single ``rating'' edge from user to item. In other words, the first connection from a user is assumed to be to an item and is assumed to have a weight that represents the user rating with higher rated items being more preferred. This construction is common in recommendation contexts where users' quantitative preferences can be gathered. The proposed algorithm is not limited to a single weighted edge and can be extended for any number of weighted edges.

	\begin{algorithm}
		\caption{Random walk meta-path generation}\label{rw}
		\begin{algorithmic}
			\Require $l \gets [ u ]$ // Initialize path with starting node: user
			\Require $m \gets$ metapath // Queue of edge types
			\Function{Generate}{l,m}  
			\If{$m \neq \{\}$}
			\State $me \gets$ \Call{pop}{m}; // Next edge type
			\State $n \gets l[1]$ //Current node
			\State $E \gets$ \Call{GetEdges}{n, me} // Get edges of type me
			\If{$me =$ \mbox{user-item}}
			\State $\langle n,j,v \rangle \gets$ \Call{WSample}{E} //weighted
			\Else
			\State $\langle n,j,v \rangle \gets$ \Call{USample}{E} //uniform
			\EndIf
			\State \Call{push}{$j, l$}; //Add node j to path
			\State \Call{MPgenerate}{l,m}
			\Else
			\State \Return $l$ 
			\EndIf
			
			\EndFunction
			
		\end{algorithmic}
	\end{algorithm}

Random walk meta-path expansion uses the process shown in Algorithm~\ref{rw}. The function \textit{Generate} takes as input a single user as the start node and a meta-path and returns a single random walk starting from the user node and guided by the meta-path. The \textit{USample} function implements uniform sampling of the edge set: an edge is chosen at random from the edges available at the current node. The \textit{WSample} function also selects a random edge from the edge set. However, the probability of choosing an edge is proportional to $e^w$, where $w$ is the edge weight. 

For example, in a movie dataset, consider the goal of generating expansions of the meta-path $user-movie-actor-movie$, for user $u$. The algorithm would proceed as follows:

\begin{enumerate}
  \item The random walker starts from $u$. Is the next edge weighted?  Yes, $user-movie$ is a weighted edge, and there are three such edges $e_1$ (weight 5), $e_2$, (weight 1), and $e_3$ (weight 3). 
  \item Function \textit{WSample} returns a weighted edge $user-movie$ in the way that the probability of selecting that edge is proportional to $e^w$, where $w$ is the edge weight. In our example, $e_1$ would be chosen with probability 87\%, $e_3$ with probability 12\%, and $e_2$ with probability 1\%. 
  \item Random walker moves from $u$ to a movie $m$ based on selected edge in step 2.
  \item Random walker is at the movie $m$. The next edge type in the meta-path is $movie-genre$. The algorithm checks if the next edge is weighted or not.
  \item The next edge $movie-genre$ is not weighted. Therefore, function \textit{USample} returns a random edge from movie $m$ and genres that $m$ belongs to.
  \item Random walker moves to genre $g$ based on selected edge in step 4. 
  \item Considering target meta-path, next edge is $movie-genre$. The $movie-genre$ an unweighted edge, therefore function \textit{USample} is again used, returning a random edge that leads to a movie $m \textprime$.  
  \item Random walker moves to movie $m \textprime$.
\end{enumerate}

Although all the meta-paths generated in this paper started from a user node and an initial weighted edge, this algorithm is sufficiently flexible to handle meta-paths where the weighted edge appears at any point in the path. For example, if an unweighted social relation of the user was available in the movie dataset, such as $user-friend$, a meta-path $user-friend-movie-actor-movie$ can be generated following this algorithm. The application of the algorithm to paths with multiple weighted edges is something we hope to explore in future work.

\section{Model Generation}

We are interested in recommendation models that make use of multiple relations simultaneously. For the work reported here, we build our recommendation models using multi-relational matrix factorization (DMF) \cite{Drumond:2014:OMF:2661829.2662052,DBLP:conf/flairs/VahedianBM16}, and incorporate relations built from the extended meta-paths described above. In multi-relational matrix factorization, one \textit{target} relation is predicted and the remaining \textit{auxiliary} relations are used as side information. For example, if the task is to recommend movies to users, the user-movie relation is the target relation and the other links between nodes such as movie-genre and movie-actor are auxiliary. We have shown that in our previous works that incorporating extended relations based on meta-paths improves the accuracy of recommender systems based on multi-relational matrix factorization \cite{DBLP:conf/recsys/VahedianBM15,DBLP:conf/flairs/VahedianBM16}. 

For the experiments described below, we compare factorized models built using our prior unweighted breadth-first meta-path expansion with those built using the sampling-based algorithm described above. In each case, we use two- and three-step meta-paths originating from the user node. We have developed in our prior work a meta-path pruning technique based on normalized information gain~\cite{something}. Pruning removes the meta-paths that contribute little to recommendation accuracy, improving both performance and training time. 

\section{Experiments and Evaluation}
For each dataset, the target relation was randomly partitioned into 80\% training and 20\% test data. Relations were generated from the training data starting with the direct relations used in the basic DMF model and adding two-step and three-step meta-paths starting from the first entity of the target relation. 

We built multi-relational factorization models for each collection of relations using the implementations of DMF made available by the authors of \cite{Drumond:2014:OMF:2661829.2662052}. \footnote{http://ismll.de/catsmf/mrFac.tar.gz} This implementation is self-contained and requires no external parameter setting other choosing an optimization criterion. We chose Bayesian Personalized Ranking as the optimization criterion (BPR-opt), as described in~\cite{Drumond:2014:OMF:2661829.2662052}. For all recommendation models, we evaluated recall and precision on recommendation lists of length one through ten.
\subsection{Datasets}
\begin{figure*}[tbh]
	\centering
	\includegraphics[width=15cm]{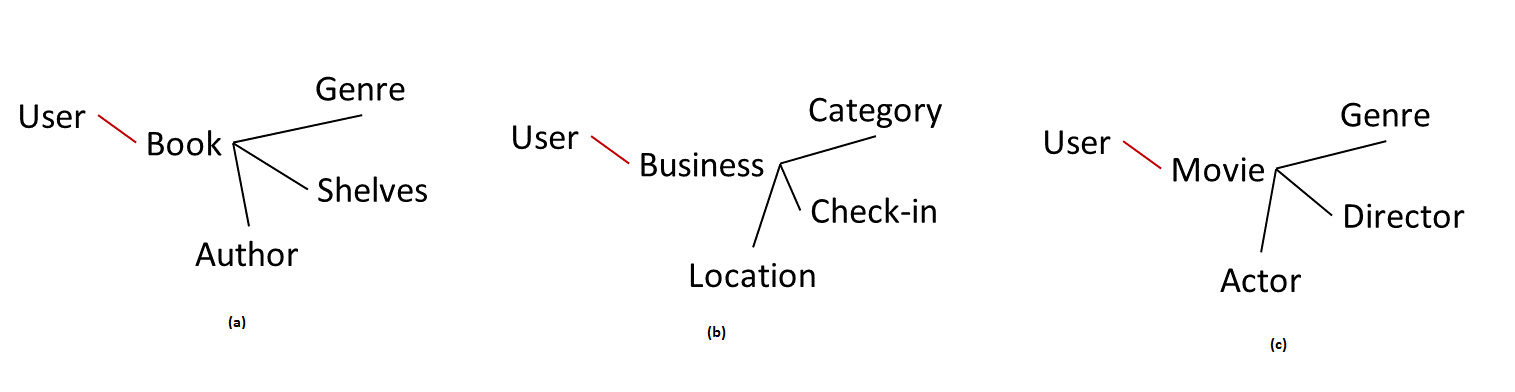}
	\caption{\label{fig:schema} Network Schema: (a) Book Crossing, (b) Yelp, (c) MovieLens}
\end{figure*}
In our experiments, we used three real world datasets in different areas. Below is a detailed explanation of these datasets.

\subsubsection*{Yelp}
This dataset\footnote{https://www.yelp.com/dataset-challenge} contains user ratings of different businesses. The node types in this network are user, business, location, category, and check-in. We performed P-core filtering as described in \cite{Gemmell:2012:RRS:2169472.2169715} to remove users and businesses with fewer than 5 ratings. The network schema of this network can be found in Figure \ref{fig:schema} (b) showing the node types and relations among them. The generated meta-paths for this network are: $user-biz-category$, $user-biz-checkIn$, $user-biz-location$, $user-biz-category-biz$, $user-biz-chekIn-biz$ and $user-biz-location-biz$. 

Table \ref{tab:yp} shows detailed statistics of the number of nodes and relations in this dataset as well as number of relations in each generated meta-path (Mp).
    
    \begin{table}[ht]
  \centering
  \caption{Yelp dataset- Number of node types and relations  }
  \label{tab:yp}
  \begin{tabular}{l|r||l|r ||l|r}
    Node & \# & Edge & \# & Mp & \#\\
    \hline
    \textbf{u}ser  & 11,937 & ub & 216,000 &ubc & 234,498\\
    \textbf{b}usiness  & 6,884 & bc & 20,409 &ubh & 1,113,600\\
    \textbf{c}ategory  & 443 & bh & 74,426 & ubl& 84,534\\
    c\textbf{h}eckIn  & 165 & bl & 6,750 &ubcb & 37,993,942\\
    \textbf{l}ocation  & 160 &  & & ubhb& 28,470,410\\
      &  &  & & ublb& 10,796,490\\
  \end{tabular}
\end{table}
    
\subsubsection*{Book Crossing}
    
Book Crossing \footnote{http://www2.informatik.uni-freiburg.de/~cziegler/BX/} contains user rating of books. Using ISBN as a key, we gathered additional book information from the GoodReads book site\footnote{https://www.goodreads.com/} including genre, tags (called ``book shelves'' in GoodReads) and authors. As in the Yelp case, a 5-core of the network was extracted so that each user and each book had a least 5 ratings. As Figure \ref{fig:schema} (a) shows, there are 5 nodes in this network: user, books, genres, shelves and authors. In this network, the $user-book$ edge is weighted in the range of 1-10. Meta-paths $user-book-genre$, $user-book-shelve$, $user-book-author$, $user-book-genre-book$, $user-book-shelve-book$, $user-book-author-book$ were generated based on random walk sampling. See Table \ref{tab:bx} for details regarding number of entities, relations and generated meta-paths. 
    
    \begin{table}[ht]
  \centering
  \caption{Book Crossing dataset: Number of node type and relations  }
  \label{tab:bx}
  \begin{tabular}{l|r||l|r ||l|r}
    Node & \# & Edge & \# & Mp & \#\\
    \hline
    \textbf{u}ser  & 7,026 & ub & 118,701 &ubg & 245,135\\
    \textbf{b}ook  & 9,432 & bg & 64,584 &ubs & 4,499,328\\
    \textbf{g}enre  & 1,689 & bs & 1,803,222 & uba& 82,627\\
    \textbf{s}helve  & 65,379 & ba & 11,599 &ubgb & 31,011,825\\
    \textbf{a}uthor  & 4,073 &  & & ubsb& 37,738,430\\
      &  &  & & ubab& 1,090,022\\
  \end{tabular}
\end{table}

\subsubsection*{MovieLens}

This dataset is an extension of well-known MovieLens 1M dataset\footnote{http://grouplens.org/datasets/movielens/1m/}. We linked the movies of MovieLens dataset with their corresponding web pages at Internet Movie Database (IMDb) and Rotten Tomatoes movie review systems. In this dataset, the weighted edge represents the user rating values for each movie. The nodes in this this network are shown in Figure \ref{fig:schema} (c). The generated meta-paths is this network were: $user-movie-genre$, $user-movie-actor$, $user-movie-director$,  $user-movie-genre-movie$, $user-movie-actor-movie$ and $user-movie-director-movie$. Details about the nodes and relations appear in Table \ref{tab:bx}.

   \begin{table}[ht]
  \centering
  \caption{MovieLens dataset: Number of node types and relations  }
  \label{c}
  \begin{tabular}{l|c||r|c ||l|r}
    Node & \# & Edge & \# & Mp & \#\\
    \hline
    \textbf{u}ser  & 2,113 & um & 855,597 &umg & 36,538\\
    \textbf{m}ovie  & 10,197 & mg & 20,809 &umd & 450,973\\
    \textbf{g}enre  & 20 & md & 10,155 & uma& 7,434,030\\
    \textbf{d}irector  & 4,060 & ma & 231,742 &umgm & 11,845,899\\
    \textbf{a}ctor  & 95,319 &  & & umdm& 2,747,252\\
      &  &  & & umam& 7,245,378\\
  \end{tabular}
\end{table}

\subsection{Comparison Algorithms}

In order to compare the accuracy of our recommender  model to graph based recommendation algorithms, we use random walk based algorithms which are also considered state of the art model in graph-based recommendation settings.

\begin{itemize}

\item \textit{ $P_{\alpha}^3$} \cite{Christoffel:2015:BWA:2792838.2800180}
The nodes in a bipartite  graph are ranked based on transition probabilities after short random walks between users and items. $P^3$ perform random walks of fixed length 3
 starting from a target user vertex. This model raises the transition probabilities to the power of a fitted parameter $\alpha$, which has been shown to improve accuracy.
 
\item \textit{$RP_{\beta}^3$} The popularity-based re-ranking model which is proposed \cite{Christoffel:2015:BWA:2792838.2800180} to compensate for the influence of  popular items in the recommendation list. 

\item \textit{$HL$} A node ranking algorithm to increases both recommendation accuracy and diversity~\cite{Christoffel:2015:BWA:2792838.2800180}. This model is a weighted linear aggregation of scores from two algorithms which are HeatS \cite{zhou2010solving} a heat diffusion across the bipartite user-item graph and ProbS \cite{zhou2010solving}, which is an item ranking method similar to $P_{\alpha}^3$.
\end{itemize}
For the  three models:{ $P_{\alpha}^3$}, {$RP_{\beta}^3$} and {$HL$}  we used the provided  code by authors \footnote{Java port: github.com/jcnewell/MyMediaLiteJava} to run the random walk based model experiments. We also used the parameter tuning suggested in \cite{Christoffel:2015:BWA:2792838.2800180} to find the best value of $\alpha$, $\beta$ and $\lambda$ for each dataset. 

In addition to these graph-oriented recommendation algorithms, we used the original DMF model restricted to only the direct network relations, to demonstrate the benefit of adding extended paths. Finally, we also used another multi-relational factorization model, CATSMF, using direct links of the network \cite{Drumond:2014:OMF:2661829.2662052}. The CATSMF model~\cite{Drumond:2014:OMF:2661829.2662052} was introduced to improve the efficiency of the DMF model by limiting the parameters needed for the auxiliary relations by coupling them together.

\section{Results and Discussion}

For each dataset, we used five-fold cross validation, computing recall and precision for each user in each fold using recommendation lists of length 1 through 10. We also calculated the normalized information gain for each meta-path generated. In the figures below, the labels DMF and CATSMF are used for the original (direct relation) factorization model, DMF-2 is the label for the extended meta-path model with two-step paths, and DMF-3 is used for the extended meta-path model with both two- and three-step paths. DMF-IG is a variant of DMF-3 where the paths with low information gain have been pruned.

\subsection{MovieLens}

\begin{figure}[tbh]
	\centering
	\includegraphics[width=8cm]{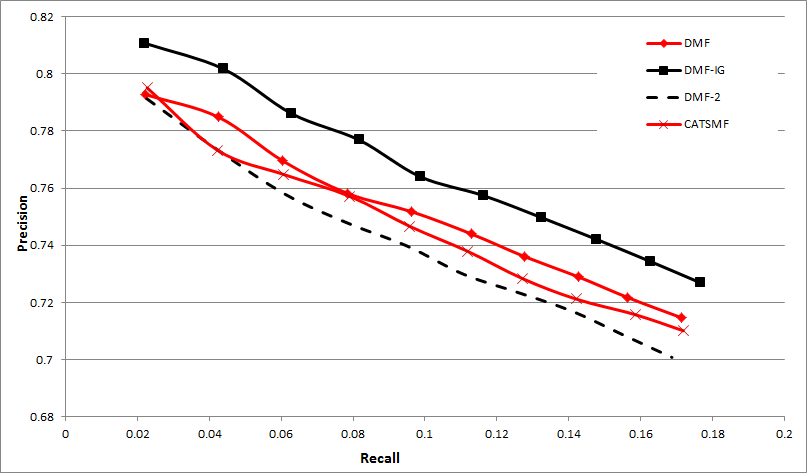}
	\caption{\label{fig:Mlresult} MovieLens dataset: Recall vs. precision for DMF models}
\end{figure}

\begin{figure}[tbh]
	\centering
	\includegraphics[width=8cm]{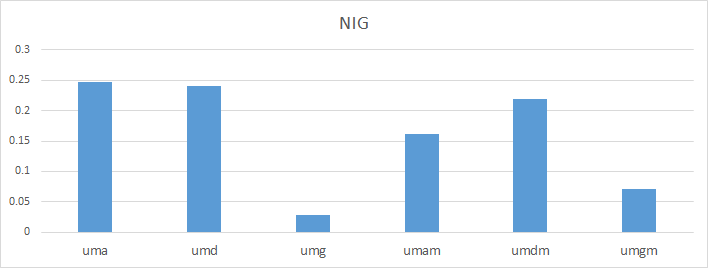}
	\caption{\label{fig:ml-nig}Normalized Information gain value for MovieLens dataset}
\end{figure}

\noindent Figure~\ref{fig:Mlresult} compares the different DMF variants, showing recall and precision curves for movie recommendation lists of length one through ten. Interestingly, DMF-2 has lower accuracy than the model using only the direct relations. Figure~\ref{fig:ml-nig} suggests a possible reason for this: the very low information gain for the genre relation, which is not surprising since there are a relatively small number of very common genre categories for movies, and knowing that a user likes one comedy, for example, is not very predictive about other comedies. On the other hand, DMF-IG, in which the less informative $user-movie-genre$ and $user-movie-genre-movie$ relations are removed, offers superior accuracy across the different list sizes. 

\begin{figure}[tbh]
	\centering
	\includegraphics[width=8cm]{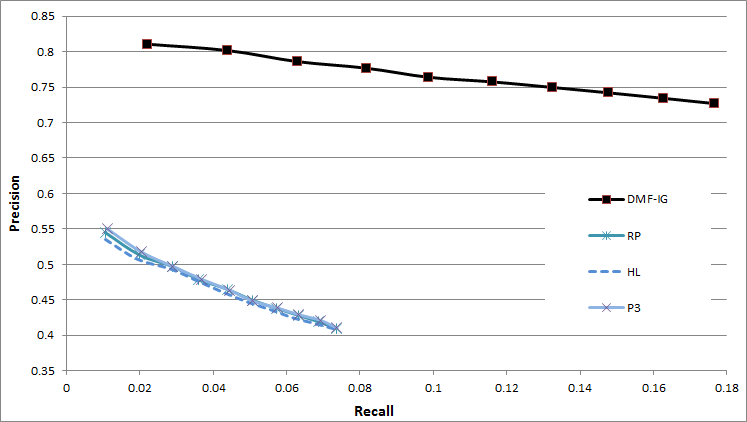}
	\caption{\label{fig:MlresultRW}MovieLens dataset: Recall vs. precision: DMF-IG vs. random-walk based recommenders}
\end{figure}

We separated out the random-walk recommenders into Figure~\ref{fig:MlresultRW} because of scale differences. As the figure shows, these algorithms have considerably worse recall/precision performance than their multi-relational counterparts.

\subsection{Book Crossing}

\begin{figure}[tbh]
\centering
	\includegraphics[width=8cm]{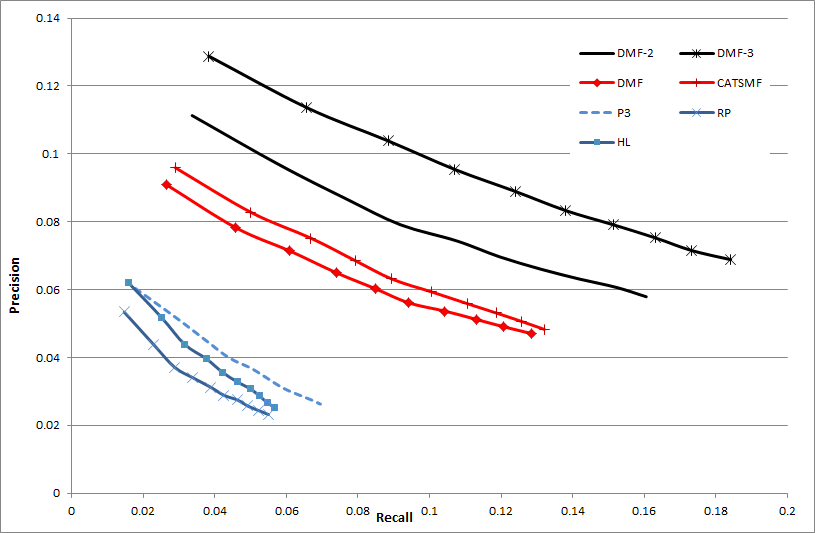}
	\caption{\label{fig:bx-rp}Recall vs. precision for Book Crossing dataset}
\end{figure}

The precision and recall curve for Book recommendation in Book Crossing dataset is shown in Figure~\ref{fig:yp-rp-mf}  for all seven algorithms. The black lines show our proposed methods, red curves represent the baseline multi-rleational matrix factorization and blue line show graph based recommendation algorithms. In the Book Crossing dataset, adding random walk based meta-paths significantly enhances the accuracy of recommendation model and the DMF-2 and DMF-3 outperform all the baseline methods. No pruned model was generated for this dataset because all of the meta-paths had good information values.

\subsection{Yelp}

\begin{figure}[tbh]
\centering
	\includegraphics[width=8cm]{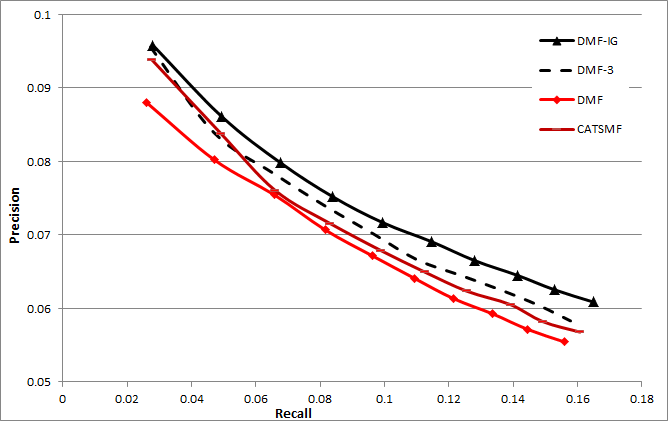}
	\caption{\label{fig:yp-rp-mf}Recall vs. precision for Yelp dataset}
\end{figure}

\begin{figure}[tbh]
	\centering
	\includegraphics[width=8cm]{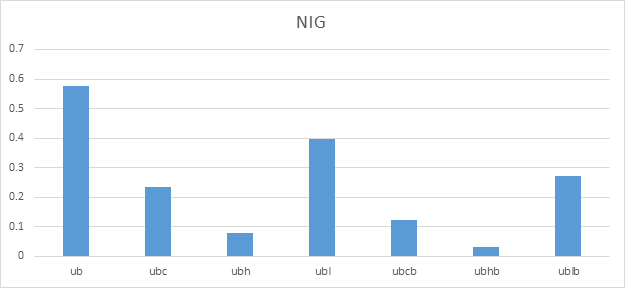}
	\caption{\label{fig:yp-nig}Normalized Information gain value for yelp dataset}
\end{figure}

Figure~\ref{fig:yp-rp-mf} depicts precision and recall results for business recommendation in Yelp network. The normalized information gain value of generated meta-paths in Yelp dataset is shown in Figure \ref{fig:yp-nig}. As Figure \ref{fig:yp-nig} shows the meta-path $user-biz-checkIn$ and $user-biz-chekIn-biz$ are the less informative paths in this network therefore, those paths were removed from the model to create DMF-IG. In Yelp datset, adding all two steps and three steps meta-paths as side information to DMF model (DMF-3) enhances the accuracy of recommendation slightly while DMF-IG shows significant improvement over both DMF and CATSMF.

\begin{figure}[tbh]
\centering
	\includegraphics[width=8cm]{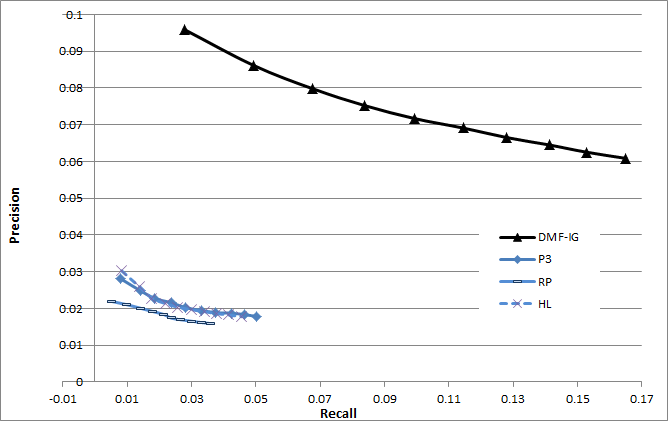}
	\caption{\label{fig:yp-rp-rw}Recall vs. precision for Yelp dataset}
\end{figure}

As with the MovieLens experiment, due to scale, we have a separate figure for the random-walk based recommenders. These algorithms show very low accuracy compared to the multi-relational alternatives.

\begin{figure}[tbh]
\centering
	\includegraphics[width=8cm]{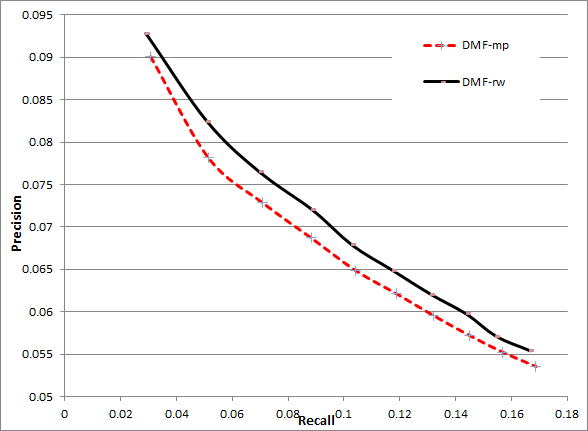}
	\caption{\label{fig:yp-r3}Recall vs. precision for Yelp dataset}
\end{figure}
In order to evaluate the ability of our model to predict highly-rated items, we performed another experiment using the Yelp dataset including just those results for items that are highly-rated by the user. The resulting metrics are referred to as US-recall and US-precision in \cite{barbieri2012balancing}. We used a threshold of 3 and only counted recall and precision ``hits'' for items with ratings above this amount. Figure~\ref{fig:yp-r3} compares recommendations generated using an unweighted relation generation technique (DMF-mp) with the algorithm described in this paper for weighted sampling (DMF-rw). 

\subsection{Efficiency}

As might be expected, random walk sampling for meta-path generation is much faster than generating the full meta-path relations. We quantify these differences in this section. All the meta-path generation experiments were executed on a Windows 7 workstation equipped with an Intel Core i7 3.40GHz  processor and 16GB RAM. The meta-path generation code was implemented in Java.

\begin{figure}[tbh]
	\centering
	\includegraphics[width=8cm]{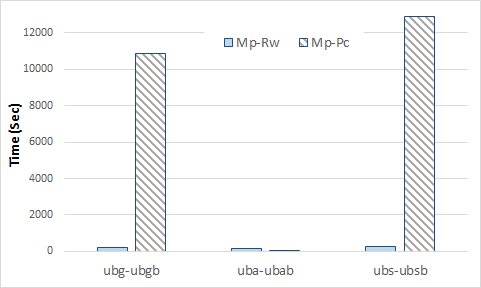}
	\caption{\label{fig:bx-tm} Book Crossing: Meta-path generation execution time}
\end{figure}

Figure \ref{fig:bx-tm} shows the time required to generate the 6 different types of meta-paths for the Book Crossing dataset using the random walk based method and the complete breadth-first method. Mp-Rw represents the execution for random walk model and MP-Pc denotes the full expansion of each meta-path. In Book Crossing dataset generating 4 of meta-paths through genre and shelve nodes by random walk sampling is much faster than original way and takes less than 5\% of the time required by the baseline technique (MP-Pc).

The difference is not notable for the meta-paths through author node, which requires about the same amount of time for both methods. The reason for this difference is that many of the books in the dataset only have a single author. The branching factor for the $book-author$ meta-path step is therefore low and the random walk and bread-first methods generate comparable numbers of paths.

\begin{figure}[tbh]
	\centering
	\includegraphics[width=8cm]{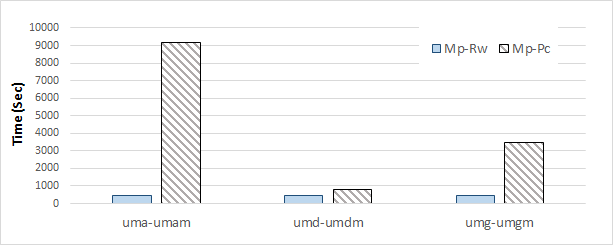}
	\caption{\label{fig:ml-tm} MovieLens dataset: Meta-path generation execution time}
\end{figure}

A similar efficiency gain is found in the MovieLens dataset. On our test machine, the random walk method takes less than 5\% of the time required by the baseline technique to generate $user-movie-actor$ and $user-movie-actor-movie$ meta-paths. The execution time is shown in Figure \ref{fig:ml-tm}. In this dataset, generating the meta-paths going through  director of movies require similar time for both approaches. Similar to authors in Book Crossing, in MovieLens, the $movie-director$ edge is for the most part an one-to-one edge while the $movie-actor$ edge might show 50 actors associated with one movie. Since meta-path generation is a major portion of the overall learning time for this system, the random sampling technique would be strongly preferred even if its accuracy were not better.

\section{Related work}
 
 Multi-relational recommender systems for heterogeneous networks have been studied extensively in recent years,  Yu et .al  proposed a method \cite{yu2014personalized,yu2013recommendation} to combine user feedback with different types of entity relationships in a collaborative filtering way. This study take advantage of  diffused  observed user implicit feedback along different meta-paths to generate possible recommendation model. The prediction model is obtained by applying matrix factorization techniques on the diffused user preferences to calculate latent representations for users and items accordingly. These latent features are combined  in a weighted way to define a global recommendation model. Furthermore, this model generates a different entity recommendation models for different user clusters  to distinguish user interests using \textit{Bayesian Ranking} optimization. These models use the non-negative matrix factorization approach to construct recommendation \cite{yu2014personalized,yu2013recommendation}. The latent features   for factorization models requires a user-item projection of user preferences. As a result, these models only use the meta-paths which start from user type node and end at an item type node, such as $user-movie-*-movie$. The advantage of our model over this work is that we can extend both item and user profiles following meta-paths and we propose a metric to measure the effectiveness of those meta-paths.

Shi et.al \cite{Shi:2015:SPB:2806416.2806528} proposed  \textit{SemRec} model, which uses weighted meta path to consider attribute values on links in information networks. SemRec uses a path based similarity to find the similar users of a active user under a given meta-path. The rating score of the target user $u$ on an item $i$ is obtained according to the rating scores of the similar users on the item rating intensity.
Based on rating prediction model, several prediction score can be made for a pair $u,i$. This model also constructs the final hybrid recommendation model as a weighted score  to integrate generated meta-paths.

In addition to network-oriented techniques, a separate thread of research has developed in multi-relational matrix factorization to make predictions for highly correlated data. Singh and Gordon~\cite{Singh:2008:RLV:1401890.1401969} proposed collective matrix factorization, as a model of pairwise relation data. This model factors each relation matrix with a generalized linear link function in which the factors of different models are tied together if an entity type is involved in more than one relation.

Coupled matrix factorization and tensor factorization can extend the multi-relational model to deal with higher arity relations as shown in~\cite{DBLP:journals/corr/abs-1105-3422}. Recent work attempts to integrate the network-based and factorization-based strains of research, using multi-relational factorization but with relations composed from network paths.

A context-dependent matrix factorization model, $HETEROMF$, proposed in  \cite{jamali2013heteromf} that considers a general latent factor for every entity type in addition to context-dependent latent factors for every context in which the entities participate. This model learns a general latent factor for every entity and transfer  matrices for every context, to convert the general latent factors into a context-dependent latent factor.
 
\section{Conclusion}	
\noindent
In the complex social networks, it is essential to make use of all the information that is available about users and items to enhance the accuracy of personalized recommender systems. Expanding extended meta-paths to generate user and item profiles is an effective way to capture additional information in a heterogeneous network. Where meta-path includes edges that represent user ratings, it makes sense to incorporate these rating values to more precisely represent user preferences. 

In this paper we explored a weighted sampling method to generate meta-paths in weighted heterogeneous networks as the basis for multi-relational recommendation models. We showed the results for three real world datasets and found that weighted sampling can be an effective basis for recommendation generation, especially when combined with information-gain-based relation pruning. Local graph methods, including other random-walk-based algorithms, were not as effective as those incorporating extended relations. We found that random sampling of edges improved recommendation accuracy while greatly reducing the computational requirements of model generation. 

In our future work, we will be exploring other multi-relational algorithms and hybrid models using weighted sampling and extend the application of weighted sampling to other types of weighted edges including the problem of combining weights when meta-path generation involves multiple weighted edges.

\section*{Acknowledgments}
This work was supported in part by the National Science Foundation under Grant No. IIS-1423368 (Multi-dimensional Recommendation in Complex Heterogeneous Networks).

\bibliographystyle{ACM-Reference-Format}
\bibliography{sigproc} 

\end{document}